\documentclass{article}
\usepackage{ijcai23}

\usepackage{times}
\usepackage{soul}
\usepackage{url}
\usepackage[hidelinks]{hyperref}
\usepackage[utf8]{inputenc}
\usepackage[small]{caption}
\usepackage{graphicx}
\usepackage{amsmath}
\usepackage{amsthm}
\usepackage{booktabs}
\usepackage{algorithm}
\usepackage{algorithmic}
\usepackage[switch]{lineno}

\graphicspath{{figures/}}

\title{From Cubes to Networks: Fast Generic Model for Synthetic Networks Generation}

\author{
Min Shaojie$^1$\And
Liu Ji$^1$
\affiliations
$^1$Chongqing University
\emails
\{alexmin, liujiboy\}@cqu.edu.cn,
}

\begin{document}

\maketitle

\begin{abstract}
    Analytical explorations on complex networks and cubes (i.e., multi-dimensional datasets) are currently two separate research fields with different strategies. To gain more insights into cube dynamics via unique network-domain methodologies and to obtain abundant synthetic networks, we need a transformation approach from cubes into associated networks. To this end, we propose FGM, a fast generic model converting cubes into interrelated networks, whereby samples are remodeled into nodes and network dynamics are guided under the concept of nearest-neighbor searching. Through comparison with previous models, we show that FGM can cost-efficiently generate networks exhibiting typical patterns more closely aligned to factual networks, such as more authentic degree distribution, power-law average nearest-neighbor degree dependency, and the influence decay phenomenon we consider vital for networks. Furthermore, we evaluate the networks that FGM generates through various cubes. Results show that FGM is resilient to input perturbations, producing networks with consistent fine properties.
\end{abstract}

\section{Introduction}

Factual networks are often sophisticated and hard to analyze. Thanks to the discovery of common network patterns and the advancement of network-generation models \cite{TRAVERS1977179,SW,BA}, network researchers can utilize synthetic networks created by models, rather than merely relying on factual networks.
Generative network models create synthetic graphs with provided inputs using sets of pre-defined regulations and equations \cite{RN1501,RN1412}. The generated graph is only applicable when it matches patterns of reality \cite{RN1501}, i.e., if we can guarantee synthetic graphs are similar to real-life networks in certain aspects, perceptions will be provided into dynamic causalities on network development.


Integration of generative network models and real-life data inherits both the advantage of system dynamic modeling provided by unique network domain methodologies and the advantage of abundant real-life data from various research domains, i.e., to predict how a specific system evolves with the help of network approaches and to create sufficient synthetic networks with reality interrelations.

However, designing such a data-driven network model can be problematic. Firstly, heterogeneity lies in data format, i.e., traditional research mainly inquiries cubes (i.e. multi-dimensional data) whereas network studies nodes and connections. Secondly and more importantly, as Wen et al argued \cite{digital_twin} that the ultimate goal of the data-driven network model is to create a synthetic network as the “Digital Twin” of reality, it is vital for the generated network to be interrelated with the original data. Achieving this requirement is yet more arduous when we attempt the network to mimic real-life patterns. Additionally, the generator should also be cost-efficient and sustained to input perturbations. 

Current solutions for such a transformation (i.e., from real-life data to networks) have limited applicability as their focuses are mainly confined to a single dataset or a specific research topic \cite{n7,n8,n9,n10,n11}. A relatively generic solution, the OSN Evolution model \cite{n12}, introduce a similarity-based method for user-network creation. Although its approach of mapping users' characteristics into node similarities provides flexibility to the model, the generation process is computationally expensive and cannot be applied to non-user-based fields.

The primary goal of this paper is to create a generative model pursuing the aforementioned objectives. To this end, we propose FGM, {\bf F}ast {\bf G}eneric Network-Generation {\bf M}odel for transformation from real-life data to interrelated networks. Real-life data for FGM input is in the form of multi-dimensional datasets (i.e., cubes). As a consequence, we refer to the input of FGM as cubes hereafter. The model converts samples in a cube into nodes with representative attributes and the network evolution is guided by those attributes and a potential-neighbor resampling strategy.

FGM is able to generate networks with low computational complexity while still reproducing real-life phenomena (e.g., properties of factual networks and the diminishing of individual's influence). Through comparison with previous works and simulations with varied cubes, we provide validation for FGM's excellent performance and robustness to perturbations.

Our methodology is highly flexible, but to ensure generated networks remain dynamic and retain desirable properties, we stipulate two integral restrictions on the input cube:
\begin{enumerate}
    \item Individuals within ought to process orderly relationships, portraying the corresponding nodes' successive sequence of arrival.
    \item Individuals within ought to retain representative values, either generated from the original cube or manually assigned.
    \item (Optional) Individuals within may have geo-graphic feature(s) specifying the positional information, which ultimately yields a location-based network.
\end{enumerate}
It is worth mentioning that those restrictions don't have to be directly satisfied by the cube. Instead, they can be calculated or inferred. We provide detailed elaborations concerning those attributes in Section \ref{sec:nodes_generation}.

\section{Related Works} \label{sec:related_works}

\paragraph{Generative Network Models} \label{sec:related_works_gnm}
Previous generative models of networks can be roughly divided into two categories based on the formation determinant of the network. The first category is {\it property-based-solely models}, in which network arrangements and connections are solely determined by the network's own properties. This type of model is invaluable for unveiling the causality of observed patterns, yet their abilities of associating with reality are still limited because no additional data is supplemented during the procedure. Examples include BA model \cite{BA}, Small-World Network \cite{SW}, Binary Relation model \cite{JA114}, Fitness-Weighted Preferential Attachment model \cite{CP1229}, etc.
The second is {\it network models with hybrid inputs}. Inputs for this category include distribution sequences, a pre-defined graph partition, samples with attributes from a dataset, etc. Hybrid inputs provide generative models with not only additional information but also the potentiality to integrate with other research fields. Representative of this type is the Configuration model \cite{conf}. Other examples include DSNG-M \cite{dsng-m}, which produce dynamic graphs from community partitioning alongside provided initial graphs, and NetGAN \cite{netgan}, taking existing graphs with hidden links as input while learning the hidden details through sequential characteristics. To a certain degree, DSNG-M and NetGAN are analogous as they both require a part of the network provided beforehand. Those models can be innovative within their specific research focuses, however, the study of network generation with seamless integration of reality is not investigated.

\paragraph{Data-Driven Network Models}
Data-driven network models explore integration of networks and factual data. For synthetic network systems with reality interrelations, Wen et al \cite{digital_twin} proposed a uniﬁed assessment criteria and argue that the ultimate goal of such an approach is creating a Digital Twin (DT) that perfectly matches reality. Unfortunately, current methodologies still fall far short of desired outcomes.
Using network methods to model reality has received increasing attention over the years \cite{JA1316,JA1013,JA1010,JA1012}. As those explorations are restricted to a specific domain, their applicability is limited. OSN Evolution model \cite{osn} is a relatively more generic approach for user-network generation but it can be computationally expensive. To our best knowledge, there are still no generic approaches for transformation from real-life data into interrelated networks.

\section{The Model}
\begin{figure}[t]
\includegraphics[width=\columnwidth]{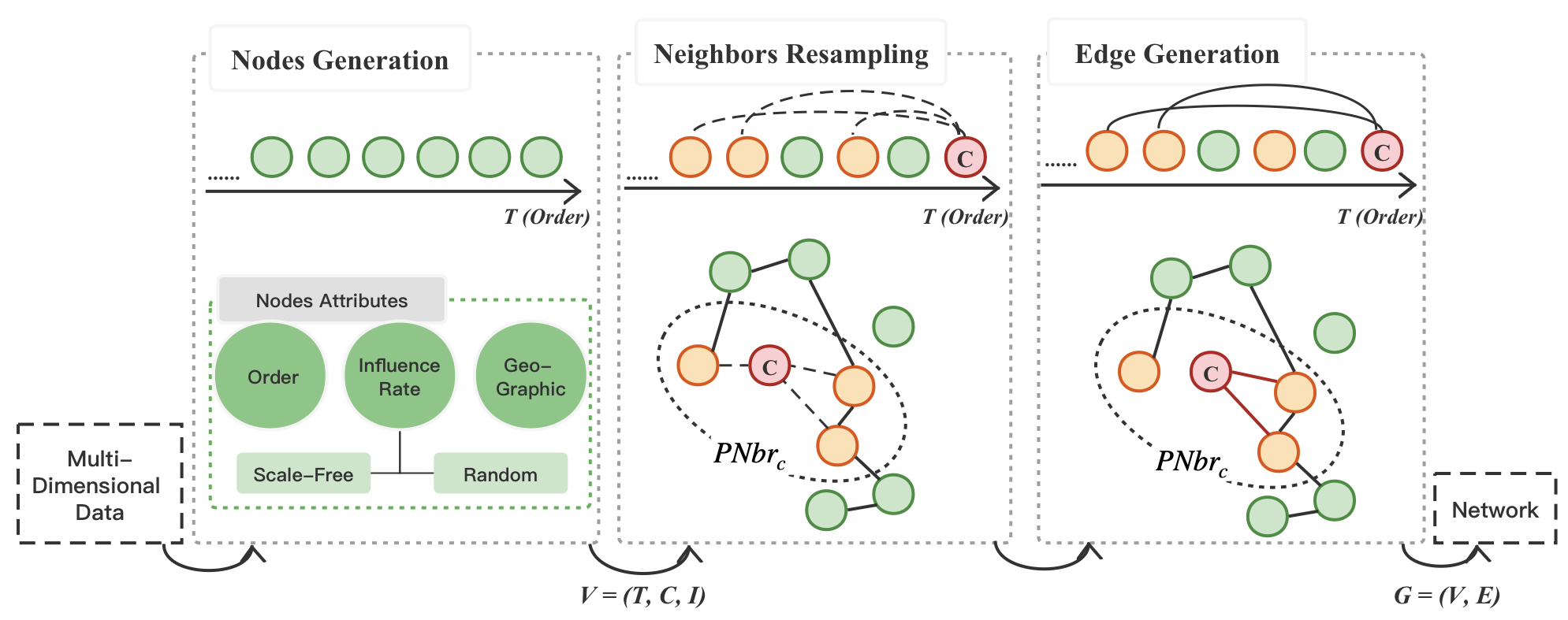}
\centering
\caption{The workflow of FGM.}
\label{fig:workflow}
\end{figure}

FGM transmutes cubes into affiliated networks. As illustrated in Figure \ref{fig:workflow}, the procedure involves three major steps: {\it nodes generation}  for converting individuals into nodes; {\it neighbors resampling} for finding each node's potential neighbors; and {\it edge generation} for settling the final connections.

\subsection{Nodes Generation} \label{sec:nodes_generation}

As a first step, we generate $N$ nodes with attributes according to the $N$ samples within the cube, denoted as $V = (T, C, I)$, where $T, C, I$ represent nodes' order, geographic and influence attributes respectively.

\paragraph{The Order Attribute}
The order attribute $T$ establishes a non-strict partial order relation“$\le$” among nodes that is reflexive, antisymmetric, and transitive. As $T$ defines the "order of arrival",  the simplest approach is employing original temporal information, nevertheless, any other intended features that satisfy the following restrictions are suitable: any $t_i, t_j, t_k\in T$ satisfies:
\begin{enumerate}
\item Reflexivity:  $t_i \le t_i$, (i.e., every element is related to itself).
\item Anti-symmetry:  if $t_i \le t_j$ and $t_j \le t_i$, then $t_i =t_i$ (i.e., no two distinct elements precede each other).
\item Transitivity: if $t_i \le t_j$ and $t_j \le t_k$, then $t_i \le t_k$.
\end{enumerate}

\paragraph{The Geographic Attribute}
The geographic attribute $C$ can have more than one dimension. It specifies the positional information among nodes for finding potential neighbors thereafter. With respect to cubes with geo-based properties such as longitude and latitude, we simply apply those properties as $C$, resulting in geo-based networks with positional nodes. For others without, we also provide an alternative strategy in Section \ref{sec:discussion}.

\begin{figure}[ht]
    \includegraphics[width=\columnwidth]{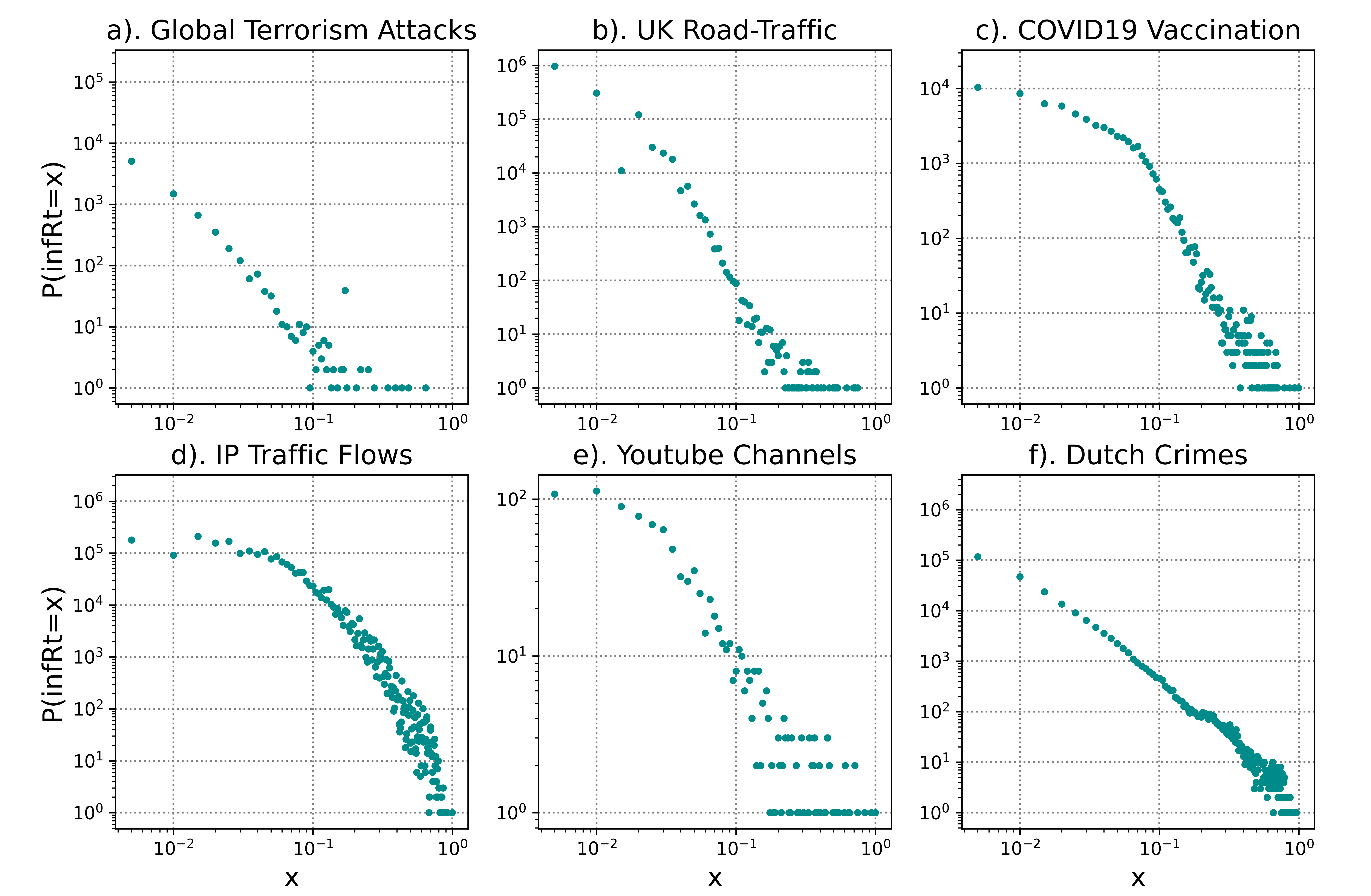}
    \centering
    \caption{Distributions (log-log) of influence attribute $I$ from various cubes.}
    \label{fig:cubes_plaws}
\end{figure}

\paragraph{The Influence Attribute}
The influence attribute $I$ indicates the node's impact on the network by dictating the number of potential edges and establishing the edge-formation probabilities during Section \ref{sec:edges_generation}.
Although $I$ can be manually assigned, our research has shown that due to the general applicability of the Pareto Principle, for a plentiful of cubes, individuals within can be portrayed with values that belong to a power-law population (examples in Figure \ref{fig:cubes_plaws} a.\cite{gtd}; b.\cite{road_data}; c.\cite{covid}, d.\cite{ipflow}, e.\cite{ytb}; f.\cite{dutch_crimes}). This transformation is accomplished via a linear function:
\begin{equation}
    infRt_{v_s} = Scale(\alpha{X_s} + \beta)
\end{equation}

where $infRt_{v_s} \in I$ is the matching node's influence attribute, $X_s$ is the original features of sample $s$, and $Scale(\cdot)$ is the min-max scaler controlling total number of edges.
As for other cases where a power-law $I$ cannot be attained, we consider them to have random $I$, and FGM is still capable of converting this type of cube into scale-free networks with desirable properties.

\subsection{Neighbors Resampling} \label{sec:neighbors_resampling}
The following step is determining candidates of final neighbors for each node (i.e., potential neighbors). The $T$ sequence yields an arrival sequence of $V$. For each $t \in T$, let $v_t$ describe the corresponding node and $G_t=(V_t, E_t)$ describe the graph. The potential neighbors of $v_t$, denoted by $\mathit{PNbr_{v_t}}$, are sampled through a nearest neighbor searching algorithm, denoted by $\Upsilon(\cdot)$:
\begin{equation} \label{eq:nnsearch}
    PNbr_{v_t} = \Upsilon_{k_t}(v_t, V_{t-1})
\end{equation}
Within $\Upsilon(\cdot)$, the number of potential neighbors (i.e., $k_t$) of node $v_t$ is determined by $v_t$'s influence attribute $infRt_{v_t}$ regardless of its geographic attribute $c_{v_t}$:
\begin{equation} \label{eq:k}
    k_t = min(n_{t-1},\ \eta*infRt_{v_{t}}^ \theta)
\end{equation}
where $n_{t-1} =\mid V_{t-1}\mid $ and $\eta, \theta \in \mathbf{R}^+$ are parameters guiding the global edge scale and the degree deviation among nodes respectively.

As we have discussed previously, $I$ is either pow-law or considered random. For the former, $\theta$ is set to 1, and for the latter, $\theta$ is generally between 3-9 to maintain power-law degree distribution. Once $k_t$ is settled, the distance measurement within $\Upsilon(\cdot)$ is obtained as:
\begin{equation} \label{eq:dist_measure}
    d(i,j) = dist_{measure}(attr_{i}, attr_{j}) 
\end{equation}

where $i,j \in V$, $dist_{measure}(\cdot)$ represents an intended measurement, and $attr$ is the weighted order and geographic attributes:
\begin{equation}
    attr_i = [\mu_tt_i,\ \mu_{c}(c_ {i,1},\ c_ {i,2}, ... , c_ {i,\lambda})]
\end{equation}
where $t_i\in T$, $c_i\in C$, $i \in V$, and $\lambda$ being the geographic dimension. Note that the order and geographic attributes (i.e., $T$ and $C$) are both put into consideration when calculating the distance. Because in FGM, it is imperative that we fully account for the impact of both order and geographic properties on node associations (the benefit will be illustrated in Section \ref{sec:influence_decay}).


As the result of {\it neighbors resampling} simply alters the recipient of connections, the only variation that can make an impact on the network's overall properties here is $k_t$. This decoupling between specific algorithms and network overall properties gives more flexibility to the model. However, with different implementations leading to varied link participants, the interrelations between generated network and original data will ultimately be affected. As a result, although various implementations can be employed here, it is advised to choose according to the data under investigation.

\subsection{Edges Generation} \label{sec:edges_generation}

After the potential neighbors (i.e., $PNbr$) are determined, edges are generated under the control of the influence attribute $I$.

For network $G_t$, let $E_{v_t, j} = 1$ indicates the existence of an undirected edge between node $v_t$ and $j$. The node $v_t$ connects to $j$ with probability:
\begin{equation}
    P(E_{v_t,j}=1)=\Gamma (infRt_{v_t}, \ infRt_{j})
\end{equation}
where $j\in PNbr_{v_t}$ and $\Gamma(\cdot)$ represents a generalized function shifting pairs of influence attributes to edge formation determinant.

\section{Evaluation of Generated Networks}

Here, we focus on patterns and measurements of synthetic FGM networks by comparing them with other generative network models. As a generic model, FGM evaluations should not be fixed with any specific cube. Hence, node attributes are obtained as follows throughout this section:
\begin{itemize}
    \item The order and geographic attributes (i.e., $T$ and $C$) are created as random samples, whereby the dimension of $C$ is set to 2;
    \item Two scenarios of the influence attribute (i.e., $I$), are evaluated. In one scenario, denoted by $\mathrm{FGM_p}$, $I$ is acquired as samples from Lomax distribution with the probability density function as $p(x)=\frac{\alpha \mu^{\alpha}}{(x+\mu)^{\alpha+1}}$, in which $\mu, \alpha$ are set to 1 and 3 respectively. In the other scenario, denoted by $\mathrm{FGM_r},$ $I$ is obtained as random samples.
\end{itemize}
Other experiment settings and regulations that are shared across this section, if not specified otherwise, are listed below:
\begin{itemize}
    \item For the scale parameters in Equation \ref{eq:k}, $\eta$ is set to 40 and $\theta$ is set to 1 for $\mathrm{FGM_p}$ and 9 for $\mathrm{FGM_r}$.
    \item Previous models under comparison include ER network, Small-World network, and the two most representative models of our model categories in Section \ref{sec:related_works_gnm} — BA network and Configuration model.
    \item The network scales under evaluation are 500, 1000, 2000, 5000, and 10000.
    \item As $T$ and $C$ are random, the nearest neighbor algorithm in Equation \ref{eq:nnsearch} and distance measurement in Equation \ref{eq:dist_measure} won't manipulate the simulation results. Nevertheless, to clarify, we employ KNN with Minkowski distance.
    \item The version of FGM won't affect what we try to convert after Section \ref{sec:other_properties}. For simplicity, we just utilize $\mathrm{FGM_p}$, directly addressing them as FGM.
\end{itemize}

\subsection{Degree Distribution}
In this experiment, we test whether synthetic networks produced by FGM portray intended degree distributions similar to reality. We use the term {\it Gnode} (giant node) referring to the few nodes with large degrees (i.e., the tail of the power-law distribution).

\begin{figure}[ht]
    \includegraphics[width=\columnwidth]{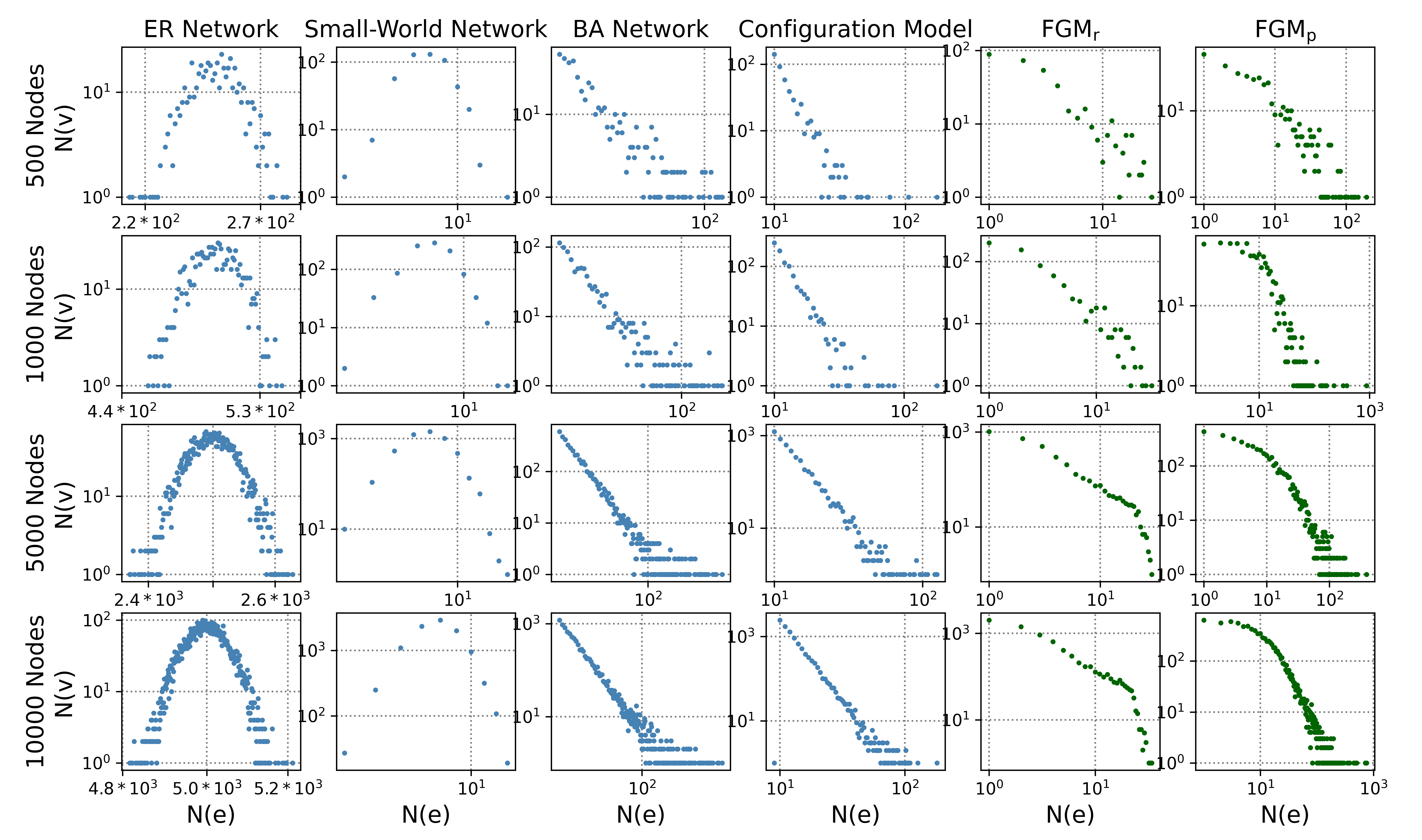}
    \centering
    \caption{Degree distributions of synthetic networks under varied network scales.}
    \label{fig:plaw_models}
\end{figure}

\begin{figure}[ht]
    \includegraphics[width=\columnwidth]{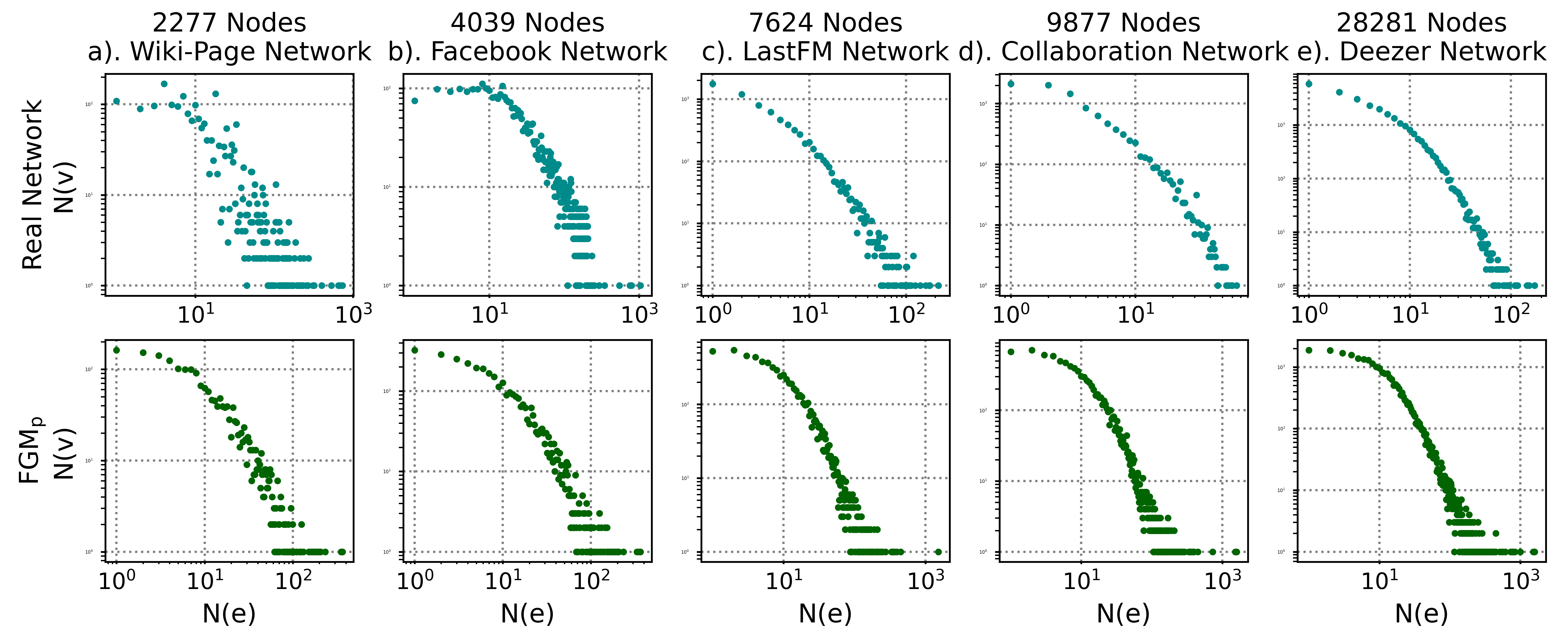}
    \centering
    \caption{Degree distributions of several factual networks and the corresponding simulations of $\mathrm{FGM_p}$.}
    \label{fig:plaw_factual_fgm}
\end{figure}

For both scenarios of FGM (see Figure \ref{fig:plaw_models}), networks are able to exhibit power-law tendencies. Compared with other models (i.e., BA network and Configuration model) where the log-log curve gradient remains constant, the curve head of FGM is flatter than the tail, particularly in $\mathrm{FGM_p}$. Since the power law often applies only for values greater than some minimum $x_{min}$  for real-life networks \cite{plaw_prove}, we believe our model better captures the nature of factual degree distribution (see Figure \ref{fig:plaw_factual_fgm}). The results also indicate that the larger the network scale is, the more obvious the pattern becomes.

Figure \ref{fig:plaw_factual_fgm} shows the degree distribution of several real-life networks (i.e., a.\cite{rnet_git_wiki}; b.\cite{rnet_fb}; c/e.\cite{rnet_deezer_lastfm}; d.\cite{rnet_physics}) and $\mathrm{FGM_p}$'s simulations with corresponding scales. Regardless of the field or scale of real-life networks, the actual distribution curve head is always flatter than the tail. This network property is correctly captured by our model while others cannot.


\subsection{Average Nearest-Neighbor Degree}

Starting here, we omit the assessment on ER networks because the randomness is too high to show observable patterns.

\begin{figure}[ht] 
    \includegraphics[width=\columnwidth]{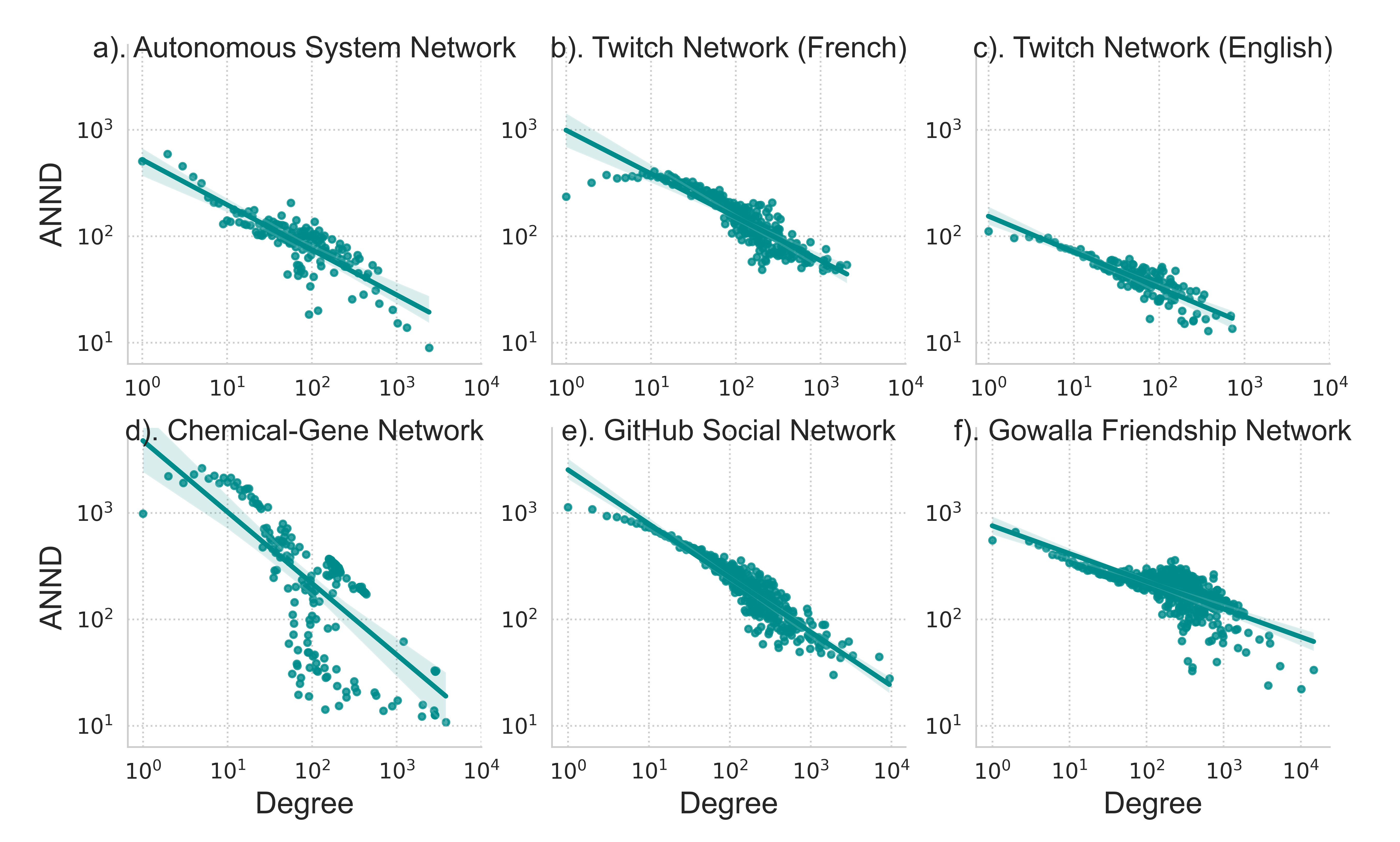}
    \centering
    \caption{ANND distributions of several factual networks.}
    \label{fig:annd_reality}
\end{figure}

Another network measurement we present in detail is the average nearest-neighbor degree (ANND) \cite{def_annd}. As a network quantization of degree-degree correlations, ANND distribution reflects the connectivity details by specifying how connected nodes with certain degrees are. When ANND was first introduced in 2001, researchers found that the Internet in 1998 exhibited a distinct power-law ANND dependence and the same patterns have been later observed to be applicable to many other factual networks \cite{def_annd,pf_annd2} (examples in Figure \ref{fig:annd_reality}, a.\cite{rnet_auto}; b/c.\cite{rnet_twitch}; d.\cite{rnet_bio}; e.\cite{rnet_git_wiki}; f.\cite{rnet_gowalla}).

\begin{figure*}[ht]
    \includegraphics[width=\textwidth]{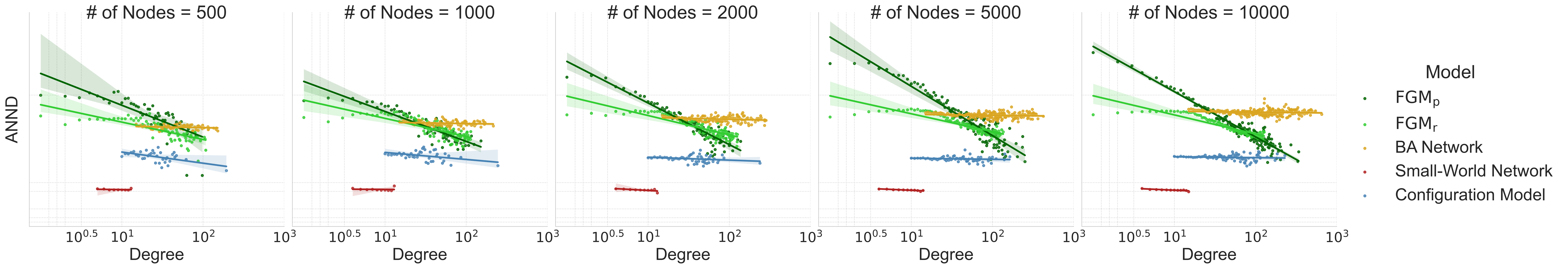}
    \centering
    \caption{ANND distributions with fitting curves of FGM and other models under varied network scales. Results show clear power-law patterns in FGM.}
    \label{fig:annd_models}
\end{figure*}

The ANND simulation results under different network scales show that both cases of FGM display a clear power-law ANND dependency while others remain relatively constant. To this concern, we also believe FGM outperforms others because the power-law ANND tendencies were also observed in reality.

\subsection{Other Network Properties} \label{sec:other_properties}

This section provides a comparison of other network properties that help assess the synthetic networks.

\begin{table*}[ht]
    \begin{tabular}{lrrlrr}
    \toprule
    Model               & \# of Nodes & \# of Edges & Parallel Edges & Clustering Coefficient & Average Path Length \\ \midrule
    Small-World Network & 500         & 2000        & No             & 0.225                  & 3.552               \\
                        & 1000        & 4000        &                & 0.221                  & 3.992               \\
                        & 2000        & 8000        &                & 0.227                  & 4.427               \\
                        & 5000        & 20000       &                & 0.227                  & 4.974               \\
                        & 10000       & 40000       &                & 0.227                  & 5.387               \\
    BA Network          & 500         & 7275        & No             & 0.126                  & 2.123               \\
                        & 1000        & 14775       &                & 0.076                  & 2.333               \\
                        & 2000        & 29775       &                & 0.050                  & 2.527               \\
                        & 5000        & 74775       &                & 0.024                  & 2.739               \\
                        & 10000       & 149775      &                & 0.014                  & 2.855               \\
    Configuration Model & 500         & 3745        & Yes            & NA                     & 2.556               \\
                        & 1000        & 7296        &                &                        & 2.783               \\
                        & 2000        & 14330       &                &                        & 3.044               \\
                        & 5000        & 36381       &                &                        & 3.389               \\
                        & 10000       & 73529       &                &                        & 3.602               \\
    $\mathrm{FGM_p}$    & 500         & 3610        & No             & 0.553                  & 2.466               \\
                        & 1000        & 7415        &                & 0.468                  & 2.968               \\
                        & 2000        & 13069       &                & 0.480                  & 3.377               \\
                        & 5000        & 39024       &                & 0.497                  & 4.323               \\
                        & 10000       & 76987       &                & 0.478                  & 5.313               \\
    $\mathrm{FGM_r}$    & 500         & 4409        & No             & 0.560                  & 2.779               \\
                        & 1000        & 9294        &                & 0.553                  & 3.319               \\
                        & 2000        & 20027       &                & 0.542                  & 3.975               \\
                        & 5000        & 53785       &                & 0.544                  & 5.234               \\
                        & 10000       & 103114      &                & 0.540                  & 6.553               \\ \bottomrule
    \end{tabular}
    \caption{Statistics of synthetic networks under different network scales (NA stands for "Not Applicable").}
    \label{tab:stats}
\end{table*}

Table \ref{tab:stats} shows that, under varied scales, FGM is the only model being able to create synthetic networks with low average length paths and high clustering coefficients while still providing network variations.

Although Small-World networks process relatively high clustering coefficients and low average path lengths, their fixed edge number severely restricts model flexibility. BA networks inherit low clustering coefficients and the coefficient seems to deteriorate while the network becomes larger. The clustering coefficient measurement for the Configuration model is not applicable due to its parallel edges, making it more troublesome to measure the network density.


\subsection{The Influence Decay} \label{sec:influence_decay}

\paragraph{Influence Decay Phenomenon} For network modeling reality, the edge-formation probability implies influence, i.e., how vital a node signifies in the network or how many links the node is able to establish. Common sense indicates that any individual's influence, even for the most important ones cannot last forever. While the system evolves, the influence of individual will diminish. This is what we address as the {\it influence decay phenomenon}.

To the network domain's concern, {\it influence decay} implies that any node's edge-forming probability ought to demise while the network evolves, i.e., any node only interacts with a vanishing fraction of previous nodes \cite{infdecay_prove}. To explain how this {\it influence decay} is uniquely reproduced in FGM, we compare FGM network with BA network from a micro-perspective: the edge-formation probability dynamics of randomly sampled {\it Gnodes}.

\begin{figure}[ht]
    \includegraphics[width=\columnwidth]{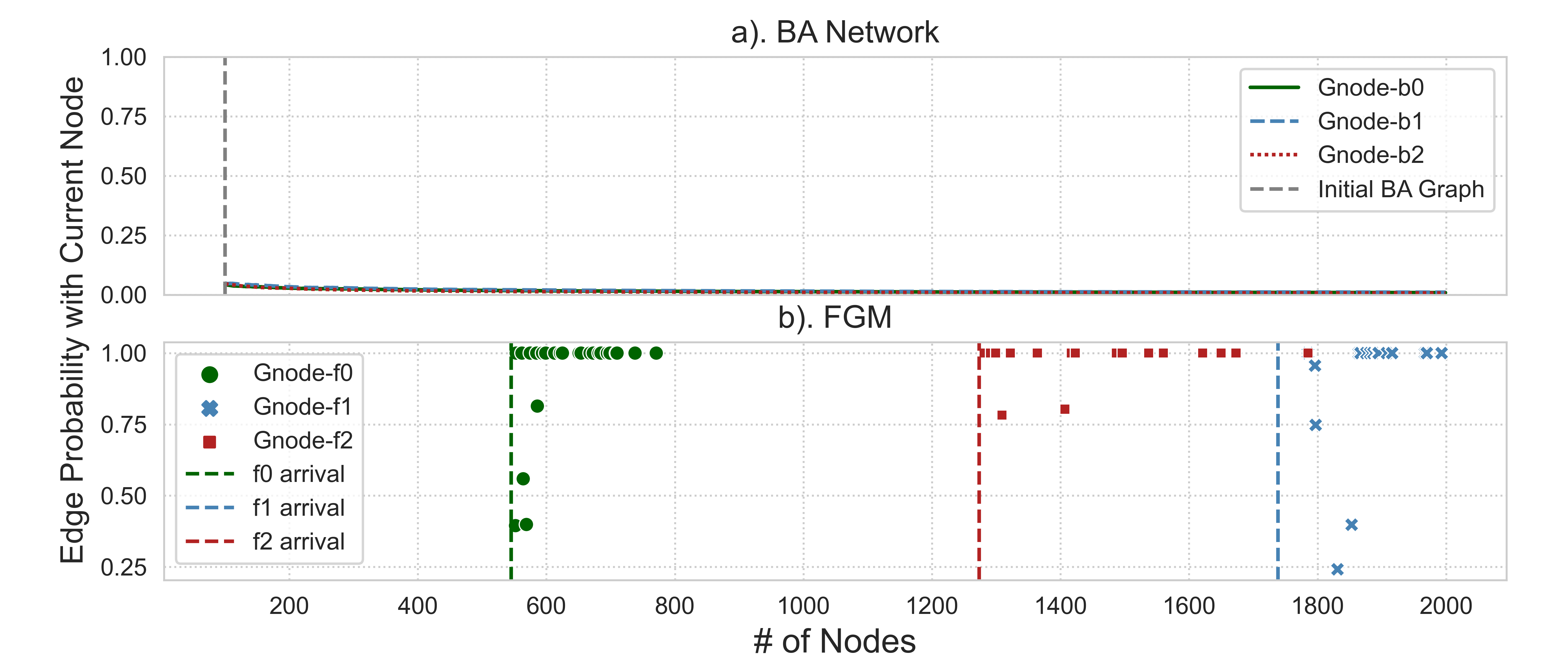}
    \centering
    \caption{Edge-forming probability dynamics of random sampled {\it Gnodes} under network of 2000 nodes.}
    \label{fig:inf_decay}
\end{figure}

For FGM, All {\it Gnodes}' probabilities dissolve at around 200 nodes after their own arrivals, satisfying the phenomenon that {\it “any individual's influence eventually demises”}. But even as the network grows 20 times bigger, those probabilities in BA network remain relatively stable, indicating sustained long-term influences.

Note that although we use the BA network as illustrations because of its profound influence, any generative model under our category {\it property-based-solely models} will potentially behave as the opposite of {\it influence decay}. Because properties inevitably grow while the network unfolds.


Although the edge-forming probabilities decay in FGM, the probability determinant $InfRt_v$ remains constant. What has changed is the less likelihood of current arrival including earlier ones in its neighbor resampling results as a result of introducing the order attribute into distance calculation.

This is an essential component of what we attempt to accomplish in FGM — the notable decayed influence with an intrinsic property remained. Because in many real-life situations, as the environment changes, individuals within have a tendency to remain in the status quo, i.e., {\it “the leopard cannot change its spots”}.

\subsection{Large Scale with High Generation Proficiency}

In this section, we demonstrate the generation time complexity of FGM by employing Locality-Sensitive Hashing (LSH) as the neighbor-searching function (i.e., $\Upsilon(\cdot$)) in Equation \ref{eq:nnsearch}.

\begin{figure}[ht]
    \includegraphics[width=\columnwidth]{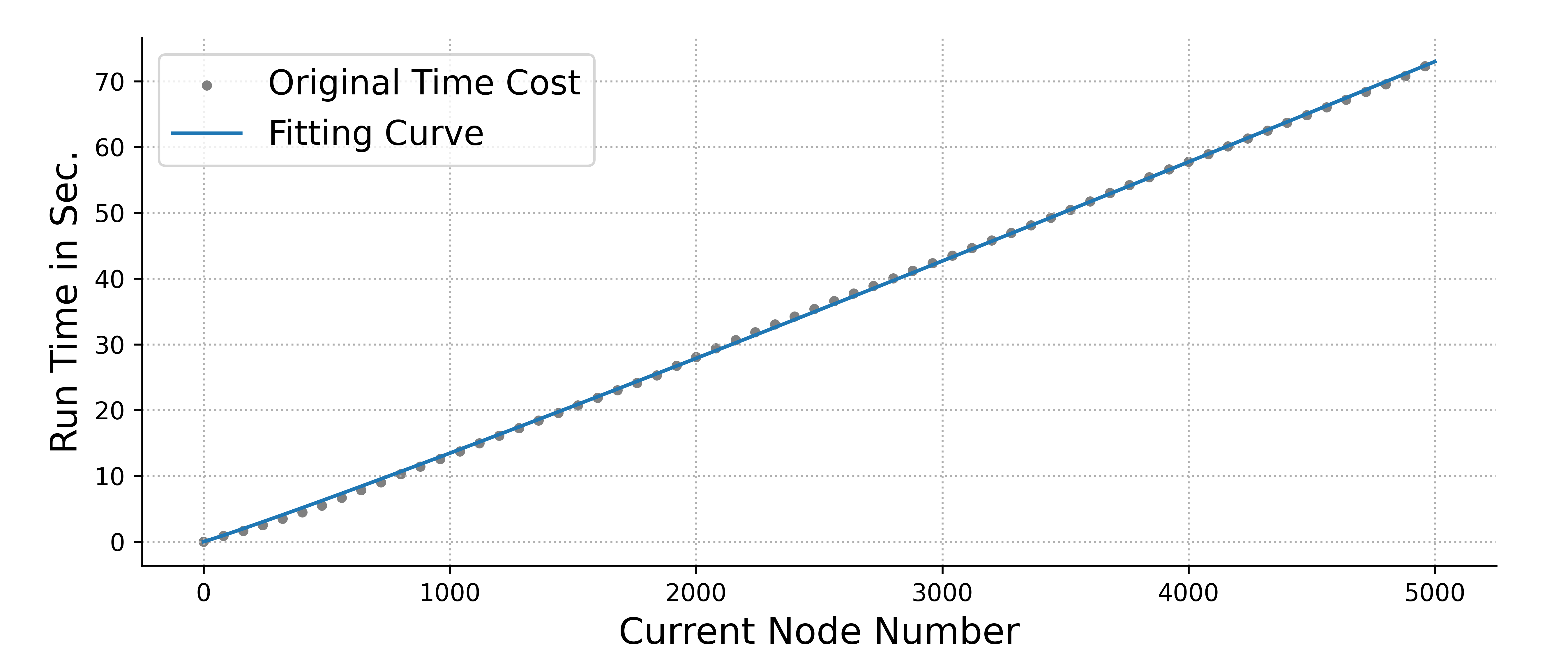}
    \centering
    \caption{Generation time of FGM with LSH and its fitting curve with original scatters in plot regularly sampled into 1/50 for legibility.}
    \label{fig:time_cost}
\end{figure}

Generation time of a 5000-node FGM network is shown in Figure \ref{fig:time_cost}. The generation time complexity of FGM is solely dependent upon the process of finding $PNbr$ owing to two factors:
\begin{enumerate}
    \item Every node only has to consider the possibility of forming edges with a small faction of existing nodes;
    \item The determinant of edge formation (i.e., $InfRt$) is constant, circumventing the necessity of recalculation when network changes.
\end{enumerate}

A constant time of finding PNbr will lead to a network generation complexity of $O(n)$, which is also the case in our demonstration of FGM with LSH's $O(1)$ query time. This makes it possible for modeling any large-scale cubes within tolerable time as dataset scale will not accumulate the computational burden.
However, it is not the case for models whose dynamics are dependent on the ever-changing network properties because of the need for recalculation. Here, we compare the time complexities of different models.

We believe the comparison of proportional time costs is much more suitable than the absolute time costs. With the very nature of network models being several pre-defined regulations, varied implementations of the same model will alter the time cost to a great extent. Still, if compared to the time cost of a small network, the proportional time costs of bigger networks will still reflect the model's tolerance of large-scale data.

Thus, every model's generation time of a 1000-node network is denoted as \{1000\} uniformly beforehand. We also include the OSN Evolution model \cite{osn} which, like FGM, models data into networks.

\begin{table}[ht]
    \centering
    \begin{tabular}{rrrrrrr}
    \toprule
    \# of Nodes         & 1000 & 2000 & 5000  & 10000  \\ \midrule
    Configuration model & 1.0  & 2.1  & 12.2  & 20.1   \\
    Small-World network & 1.0  & 4.0  & 26.2  & 96.4  \\
    BA network          & 1.0  & 6.2  & 81.6  & 606.7  \\
    OSN Evolution model & 1.0  & 9.2  & 130.5 & 1112.7 \\
    FGM                 & 1.0  & 4.1  & 25.5  & 103.8  \\ \bottomrule
    \end{tabular}
    \caption{The proportional time cost of models with unit as \{1000\}.}
    \label{tab:ptimes}
\end{table}

Table \ref{tab:ptimes} shows that within the field of scale-free networks, the Configuration model, BA model \cite{atwood_efficient_2015}, and FGM process linear generation time complexities. As for the other data-driven approach, the OSN Evolution model becomes extremely time-consuming as the scale gets bigger.

\section{Cube Experimentations}

\subsection{Properties of Cube-Networks} \label{sec:cubenets}

\begin{table*}[ht]
    \centering
    \begin{tabular}{lrrrrr}
    \toprule
    Cubes                    & \# of Nodes & \# of Edges & Clustering Coefficient & Avg. Path Length & Time Cost (Sec.) \\ \midrule
    Youtube Channels         & 1000        & 9588        & 0.458                  & 3.934            & less than 1      \\
    COVID19 Vaccination      & 86512       & 711170      & 0.375                  & 4.311            & 6                \\
    Global Terrorism Attacks & 190036      & 1036035     & 0.245                  & 5.756            & 10               \\
    UK Road-Traffic          & 2047256     & 12703808    & 0.282                  & 3.501            & 117              \\
    Dutch Crimes             & 2596254     & 14435180    & 0.251                  & 5.221            & 133              \\
    IP Traffic Flows         & 3577296     & 28797789    & 0.392                  & 3.822            & 230              \\ \bottomrule
    \end{tabular}
    \caption{Statistics of synthetic networks derived from six cubes of varied domains}
    \label{tab:cubes}
\end{table*}

In this section, we provide evaluations of FGM from a macro-perspective analysis, i.e., properties of interrelated networks from various cubes.
We convert previously illustrated cubes in Figure \ref{fig:cubes_plaws} to interrelated networks and further investigate whether synthetic networks can reproduce typical patterns as real-life networks while, in the meantime, reflecting patterns of original cubes.

\begin{figure}[ht]
    \includegraphics[width=\columnwidth]{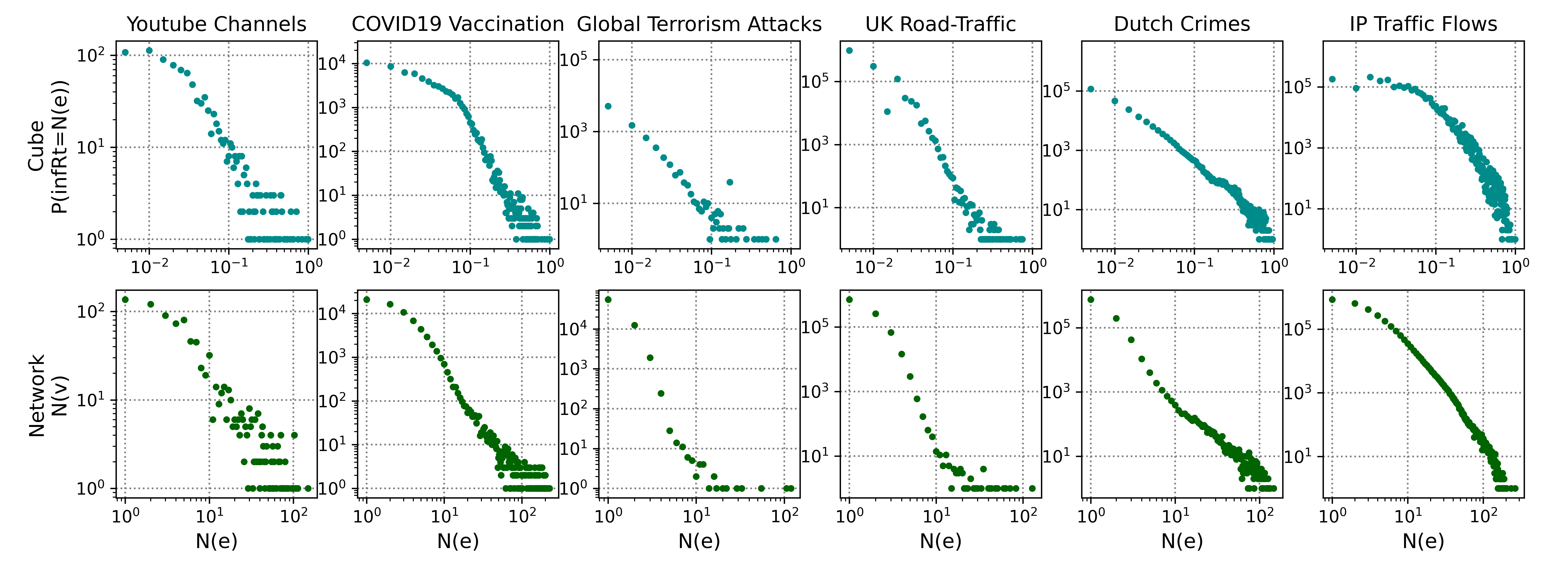}
    \centering
    \caption{Cubes $InfRt$ distributions alongside degree distributions of corresponding networks FGM generates.}
    \label{fig:cube_netwotks_plaw}
\end{figure}

From Table \ref{tab:cubes}, it is clear that, regardless of cube sizes, FGM synthetic networks can maintain high clustering coefficients with low average path lengths as factual networks. Figure \ref{fig:cube_netwotks_plaw} shows the degree distributions exhibit power-law patterns. Notably, when examined in conjunction with the $I$ distribution of the original cubes, these degree distributions also accurately capture the uniqueness of the original patterns, that is, the manner in which the data points are distributed in the plot.

Now, from a macro-perspective analysis, we believe that FGM can generate synthetic networks with interrelations to reality, where common properties of factual networks are reproduced and original data patterns are preserved to a certain degree.

\subsection{Decayed Influence of 9-11} \label{sec:911}

Here, we present the evaluations of FGM from a micro-perspective analysis, i.e., how the {\it Gnode} (i.e., 9-11 attack) in Global Terrorism Database (GTD) \cite{gtd} reflects the prescribed {\it influence decay phenomenon} (see Section \ref{sec:influence_decay}). To elaborate, we compare the edge dynamics of 9-11 attack in FGM and BA network of modeling GTD between 2001 and 2005.


\begin{figure}[ht]
    \includegraphics[width=\columnwidth]{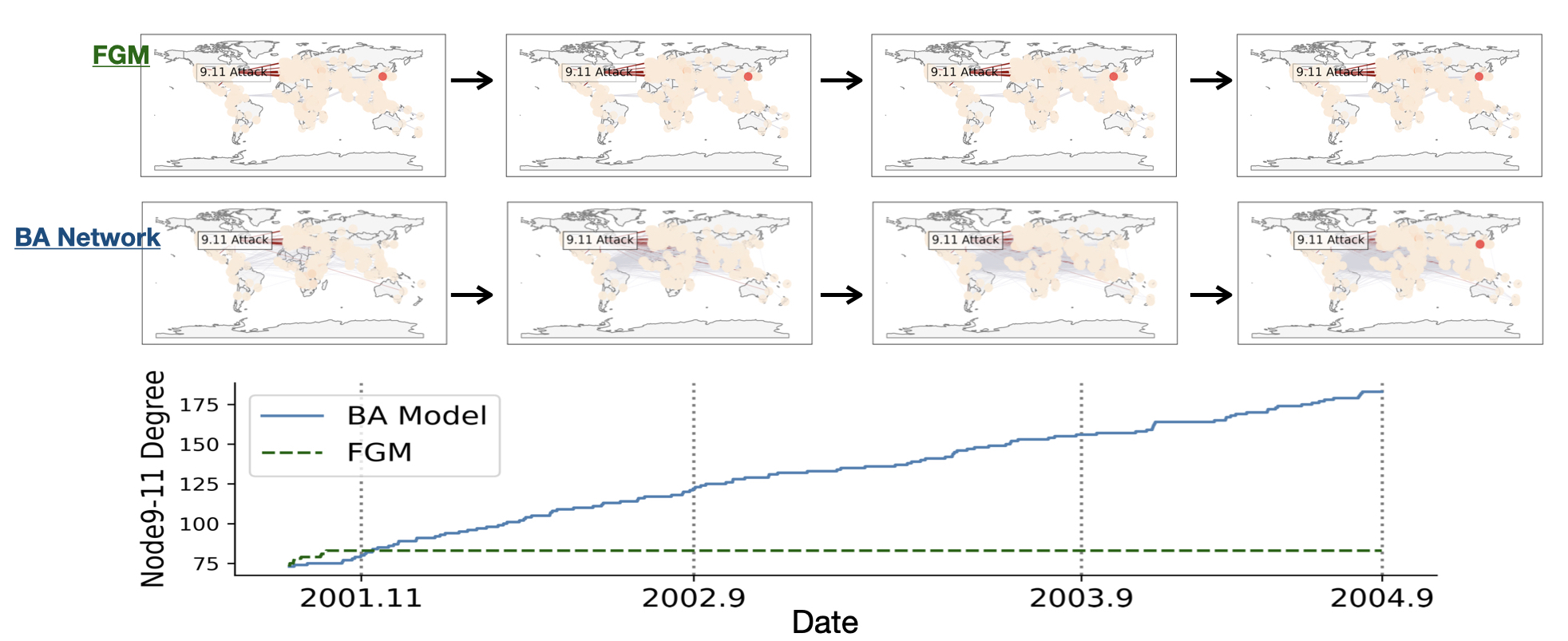}
    \centering
    \caption{GTD (2001-2005) modeling of FGM and BA networkt (line chart is horizontally aligned with network plots).}
    \label{fig:911_overall}
\end{figure}

Figure \ref{fig:911_overall} shows that, in FGM {\it Gnode 9-11}'s degree remained stationary after around 3 months whereas the edge dynamic in BA network kept evolving throughout the simulation. 
That is to say, the 9-11 influence on other attacks in FGM eventually decays despite its enormous impact at the time, whereby, in property-based model like BA Network, the impact keeps growing. To make things worse, 9-11 is forever the biggest node for each new arrivals no matter how the world evolves. Nowadays clearly more recent events have bigger impacts.

From a micro-perspective analysis, we also believe FGM reflects factual phenomenon to a centain degree where previous works cannot.

\section{Discussion} \label{sec:discussion}
During the process of neighbors resampling (see Section \ref{sec:neighbors_resampling}), we obtained potential neighbors (i.e., $PNbr$) through the order and geographic attributes (i.e., $T$ and $C$). However, we believe the acquirement of $PNbr$ (i.e. $\Upsilon(\cdot)$ in Equation \ref{eq:nnsearch}) doesn't have to utilize node attributes at all.
An alternative implementation of $\Upsilon(\cdot)$ can take the original dataset's features as input and directly output PNbr through machine learning algorithms such as Bayesian methods, Clustering, or any others that produce classification results. The direct attainment of $PNbr$ introduce better flexibility to FGM as the geographic attribute $C$ is no longer required, i.e., FGM becomes geo-free allowing for geographic irrelevance data to be modeled into non-locale-based networks.

This alternative provides yet another benefit besides more flexibility: with $PNbr$ acquired beforehand, FGM's network generation time will be fixed to O(n) regardless of model's implementation details.

\section{Conclusion}
In this paper, we answer the question “Can we efficiently model cubes from varied domains to interrelated networks?” With the proposed FGM, we believe we reach a positive answer to this question.
Extensive experiments show the generated networks can both reproduce patterns shared by factual networks and perserve the uniqueness of original data features from both marco-perspective and micro-perspective. Moreover, FGM's decoupling from specific algorithm and linear generation time provide the approach with noble flexibility and efficiency.

FGM opens doors to utilizing network-unique methodologies for systematic dynamic analysis of cubes and to generating abundant synthetic networks for future network research.

\bibliographystyle{named}
\bibliography{ijcai23}

\end{document}